\documentclass[showpacs,prl,amsmath,amssymb,superscriptaddress,nobibnotes, twocolumn, preprintnumbers]{revtex4-1}

\usepackage{graphicx}
\usepackage{endnotes}
\usepackage{bm}
\usepackage{epsfig}
\usepackage{amsmath}
\usepackage{color}

\newcommand{\ie}{{\it i.e.}}

\begin{document}
\title{Quasilinear quantum magnetoresistance in pressure-induced nonsymmorphic superconductor CrAs}
\author{Q.~Niu}
\author{W.~C.~Yu}
\author{K.~Y.~Yip}
\author{Z.~L.~Lim}
\affiliation{Department of Physics, The Chinese University of Hong Kong, Shatin, New Territories, Hong Kong, China}

\author{H.~Kotegawa}
\author{E.~Matsuoka}
\author{H.~Sugawara}
\author{H.~Tou}
\affiliation{Department of Physics, Kobe University, Kobe 658-8530, Japan}

\author{Y.~Yanase}
\email{yanase@scphys.kyoto-u.ac.jp\\ skgoh@phy.cuhk.edu.hk}
\affiliation{Department of Physics, Kyoto University, Kyoto 606-8502, Japan}

\author{Swee~K.~Goh}
\email{yanase@scphys.kyoto-u.ac.jp\\ skgoh@phy.cuhk.edu.hk}
\affiliation{Department of Physics, The Chinese University of Hong Kong, Shatin, New Territories, Hong Kong, China}
\date{September 6, 2016}


\begin{abstract}
{\bf
In conventional metals, modification of electron trajectories under magnetic field gives rise to a magnetoresistance that varies quadratically at low field, followed by a saturation at high field for closed orbits on the Fermi surface. Deviations from the conventional behaviour, {\it e.g.} the observation of a linear magnetoresistance, or a non-saturating magnetoresistance, have been attributed to exotic electron scattering mechanisms. Recently, linear magnetoresistance has been observed in many Dirac materials, in which the electron-electron correlation is relatively weak. The strongly correlated helimagnet CrAs undergoes a quantum phase transition to a nonmagnetic superconductor under pressure. Near the magnetic instability, we observe a large and non-saturating quasilinear magnetoresistance from the upper critical field to 14~T  at low temperatures. We show that the quasilinear magnetoresistance arises from an intricate interplay between a nontrivial band crossing protected by nonsymmorphic crystal symmetry and strong magnetic fluctuations.\\
}


\end{abstract}



\maketitle

Linear magnetoresistance (MR) is a fascinating material property which can be harnessed for magnetic field sensing owing to its simple field-to-resistance conversion and its equal sensitivity across the entire field range. As such, the mechanisms underpining the linear MR continue to attract interest. 
Linear MR is most commonly observed in systems with a relatively weak electron-electron correlation and comparatively low carrier density, such as doped semiconductors \cite{Xu1997, Parish2003, Hu2008} and topological semimetals \cite{Feng2015, Liang2015, Narayanan2015,Novak2015,Feng2015}. 
The observation of a linear MR in strongly correlated electron systems has been a rare occurence. A notable case of a striking linear MR is found in BaFe$_2$(As$_{1-x}$P$_x$)$_2$ near the antiferromagnetic quantum critical point \cite{Hayes2016}, where non-Fermi liquid behaviour in the temperature ($T$) dependence of the resistivity ($\rho$) has also been observed \cite{Kasahara2010}, leading to a phenomenological model to describe the electrical transport with a scattering rate that is proportional to the quadrature sum of magnetic field and temperature \cite{Hayes2016}. To develop the concept further, it is important to study other strongly correlated electron systems near magnetic instability to expand the material base for detailed in-field investigations.

CrAs undergoes a strong first-order transition at $T_N=265$~K into an antiferromagnetic state, in which the spins assume a double helical structure (winding axis $\parallel$ $c$-axis) \cite{Motizukibook,Watanabe1969}. This antiferromagnetic state can be completely suppressed by applying 7~kbar \cite{Kotegawa2014, Wu2014}, or by substituting 5\% of phosphorous for arsenic, \ie\ CrAs$_{0.95}$P$_{0.05}$ \cite{Kanaya2004}. The magnetic behaviour of phosphorous substituted samples is similar to that of CrAs under high pressure, and hence phosphorous acts as the chemical pressure analogous to other systems such as BaFe$_2$(As$_{1-x}$P$_x$)$_2$ \cite{Klintberg2010}. Concomitant with the suppression of the helimagnetic state, superconductivity can be induced  in CrAs by pressure. The superconducting transition temperature ($T_c$) takes a dome-shaped pressure dependence, with a maximum $T_c\approx2.17$~K at 10~kbar \cite{Kotegawa2014, Wu2014}. The temperature-pressure ($T$--$p$) phase diagram constructed is reminiscent of those constructed for many quantum critical systems, {\it e.g.} refs. \cite{Mathur1998, Gegenwart2008, Paglione2010, Klintberg2012, Goh2015}, except that the magnetic transition at ambient pressure is strongly first-order, although the first-order signature is significantly weakened at elevated pressures \cite{Kotegawa2014, Wu2014}. Nevertheless, nuclear quadrupole resonance detected substantial magnetic fluctuations in the paramagnetic state \cite{Kotegawa2015}. In the phosphorous substituted series, superconductivity has not been observed due to the strong disorder introduced by phosphorous substitution, which is consistent with the unconventional nature of superconductivity \cite{Kotegawa2015}.

\begin{figure*}[!t]\centering
      \resizebox{15cm}{!}{
              \includegraphics{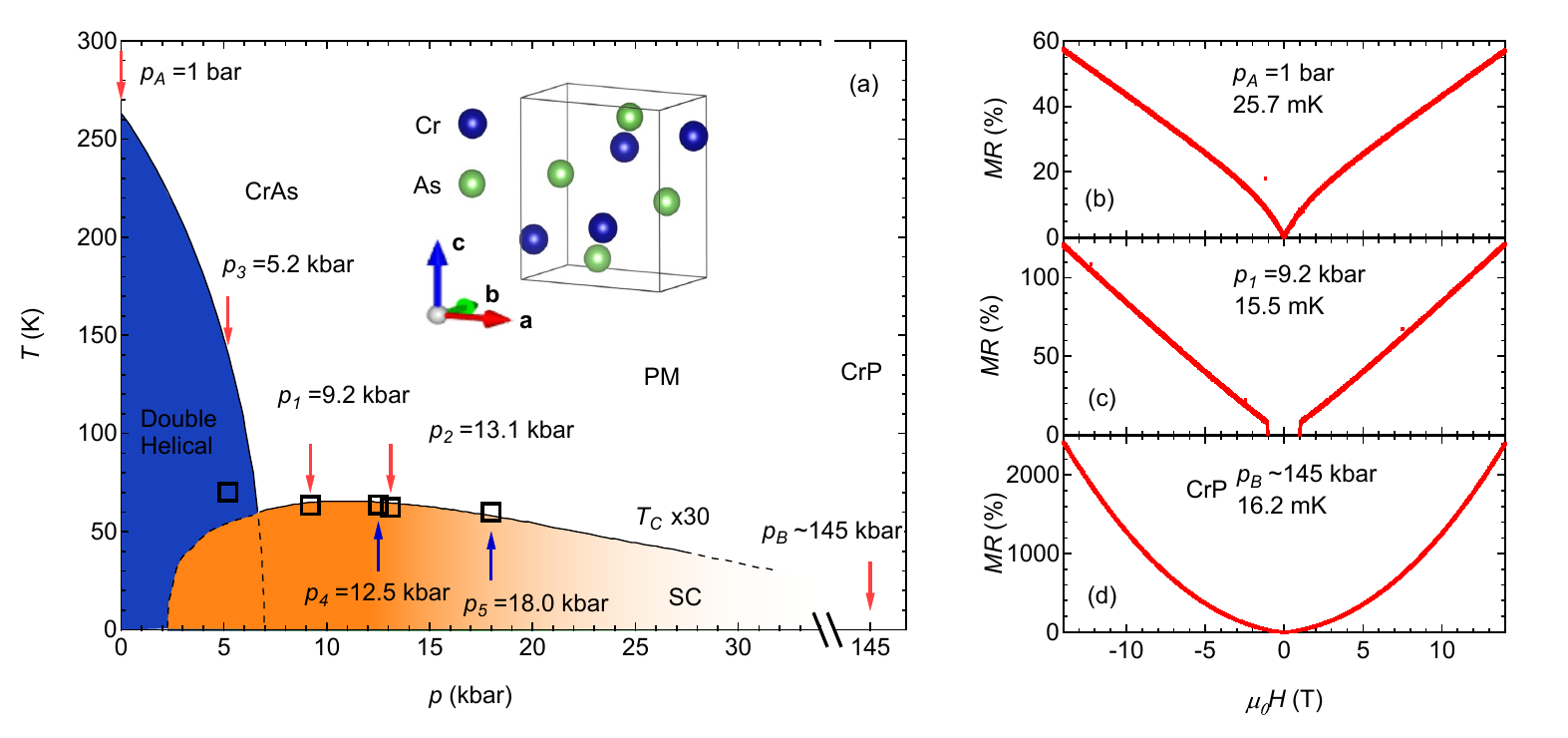}}                				
              \caption{\label{fig1} {\bf Transverse magnetoresistance in pressure-tuned CrAs. a,} Schematic temperature-pressure phase diagram showing the helimagnetic region, the superconducting region (SC) and the paramagnetic region (PM), constructed using the data in Ref. \cite{Kotegawa2014}. The area bounded by dashed lines indicates the region with coexistence of magnetism and superconductivity. The open squares denote the experimental data of the current work, which are consistent with previously published data \cite{Kotegawa2014, Wu2014}. The position of CrP is determined by taking into account the chemical pressure effect of P on CrAs. The crystal structure of CrAs, prepared using VESTA \cite{VESTA}, is shown in the inset. {\bf b,c,d,} Low-temperature magnetoresistance at $p_A$, $p_1$, and $p_B$. For the data collected in the superconducting state, the magnetoresistance is calculated using a zero field resistivity $\rho(0)$ estimated from a smooth extrapolation of the data above $H_{c2}$.}
\end{figure*}

We have measured the MR down to 14~mK in representative parts of the $T$--$p$ phase diagram (Fig. \ref{fig1}). The space group of CrAs, $Pnma$, includes nonsymmorphic glide and screw symmetries, which involve half translation. We will show that the nonsymmorphic symmetry of the crystal structure plays an essential role on the realization of a particular band structure, leading to a nontrivial band crossing on the Brillouin zone face protected by nonsymmorphic symmetry. At 14~T, the MRs are positive, sizeable and non-saturating. Furthermore, distinct quasilinear MRs are observed in the vicinity of the magnetic instability, which we will demonstrate to be a result of an intricate interplay between having this particular band structure and the proximity to magnetic instability.


\noindent{\bf $T$--$p$ phase diagram and MR}\\
Fig.~\ref{fig1}a shows the temperature-pressure phase diagram of CrAs. Three crystals were used for this study, namely CrAs at ambient pressure ($p_A$), CrP at ambient pressure ($p_B$), and CrAs under pressure ($p_i$'s with $i$ indicates the sequence of the measurement). Both CrAs and CrP crystallize in an orthorhombic structure with space group $Pnma$ (inset to Fig. ~\ref{fig1}a)\cite{Motizukibook}. Using the $x$-dependence of the unit cell volume in CrAs$_{1-x}$P$_x$ \cite{Yu2015}, we estimate that $\Delta x=1$ corresponds to $\Delta p\approx145$~kbar. Therefore, we place CrP at the location of $p=145$~kbar relative to CrAs. CrP has an excellent purity with a residual resistance ratio (RRR) of 455, in stark contrast to members of phosphorous substituted CrAs. For example, RRR$\approx$~5 for CrAs$_{0.95}$P$_{0.05}$. On the other hand, the RRR of the pressurised CrAs is far larger, {\it e.g.} RRR=194 at $p_2$, provided that care is taken not to thermally cycle the crystal through the first-order line (see below). Figs.~\ref{fig1}b--\ref{fig1}d show the MR, defined as MR$=\frac{\rho(H)-\rho(0)}{\rho(0)} \times 100\%$, at the lowest attainable temperature of each run in the helimagnetic state ($p_A$), nonmagnetic superconducting state ($p_1=9.2$~kbar) and the paramagnetic non-superconducting state ($p_B$), respectively. All traces exhibit a large and non-saturating MR at 14~T. In the helimagnetic state ($p_A$), the field dependence of the MR is sublinear at low field with a seemingly absence of the $H^2$ region. At $p_1$, the MR is quasilinear from the upper critical field ($H_{c2}$) to 14~T. Finally, at $p_B$, the MR is conventional and follows a $H^2$ variation from 0~T to 14~T. The CrAs system thus provides a useful platform for accessing different magnetoresistance behaviours by tuning the unit cell volume. All these traces are clearly an even function of $H$, indicating a negligible contamination from the Hall component. Therefore, the data presented here are truly the transverse MR and hence only the positive $H$ portion of the data are plotted in the following.

\begin{figure*}[!t]\centering
       \resizebox{15cm}{!}{
              \includegraphics{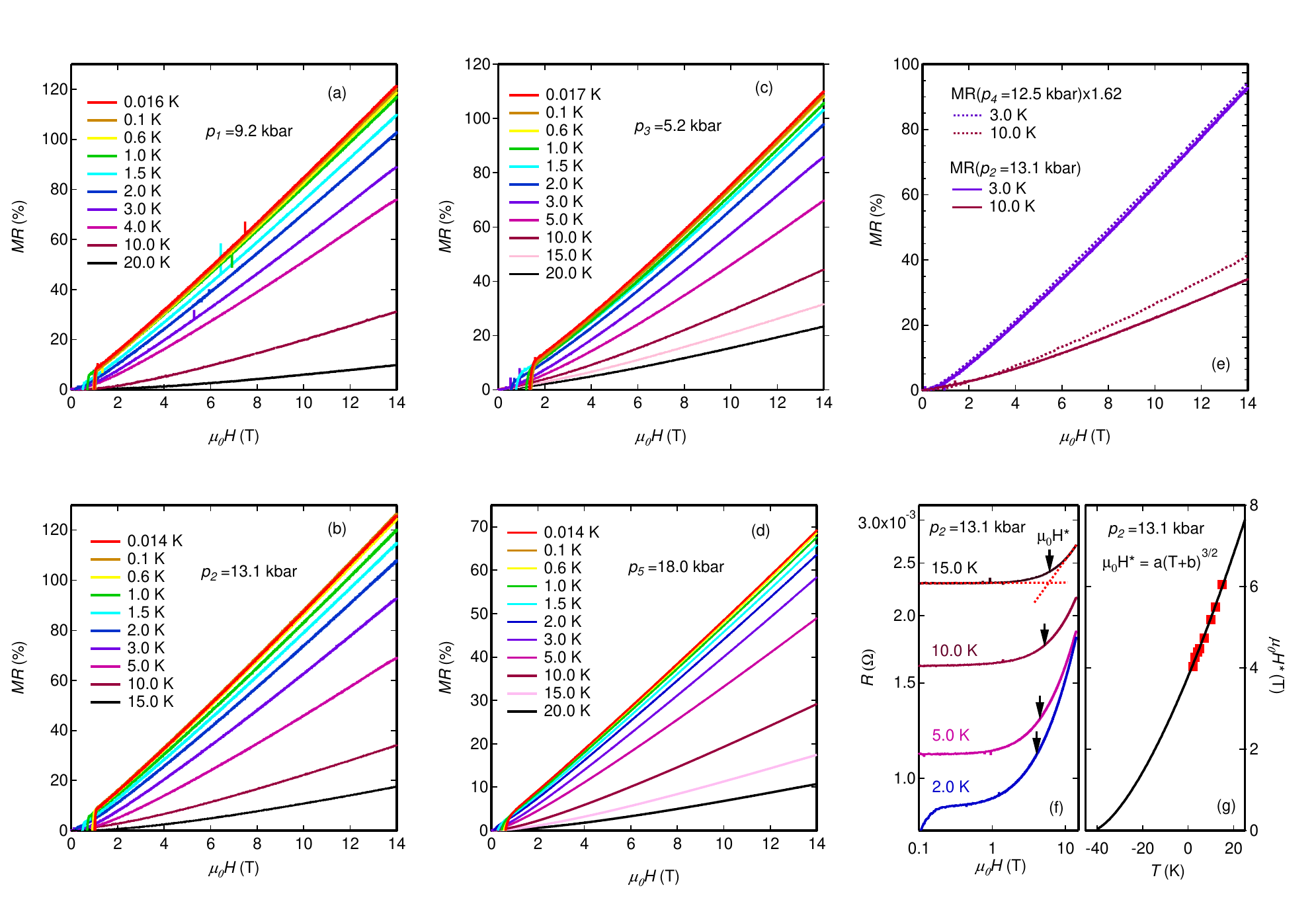}}                				
              \caption{\label{fig2} {\bf Magnetoresistance of CrAs under pressure. a,b,c,d,} Magnetoresistance of CrAs over a wide temperature range at $p_1$, $p_2$, $p_3$, and $p_5$. {\bf e,} Magnetoresistance at $p_2$ (solid lines) and $p_4$ (dashed lines). The traces at $p_4$ are multiply by 1.62, which is the ratio RRR($p_2$)/RRR($p_4$). {\bf f,} The field dependence of the resistance on logarithmic scales, showing the determination of the crossing over field $\mu_0 H^*$. {\bf g,} The temperature dependence of $\mu_0 H^*$ (squares), which can be described by $\mu_0 H^*=a(T+b)^{3/2}$ (solid line).
            }
\end{figure*}
Figs. \ref{fig2}a -- d shows the MR collected over a wide temperature range at pressures where the groundstate is superconducting. At these pressures, the MR is quasilinear at the lowest temperature, and it reaches at least 70\% at 14~T. For $p_1$ and $p_2$, the single crystal was directly pressurised to the required pressure values at room temperature before cooling. MR(14~T) at the lowest temperature for $p_1$ and $p_2$ are larger than 110\%, and it decreases with increasing temperature. In order to investigate the possibility of a disorder-induced linear MR behaviour, we take advantage of the strong magnetostriction effect associated with the first-order transition into the helimagnetic state \cite{Motizukibook}. Empirically, it is known that upon warming up across the first-order line, the crystal can crack. Therefore, we reduced the pressure to 5.2~kbar ($p_3$) and performed a thermal cycle, where the first-order transition exists but much weaker than the ambient pressure case ($p_A$) \cite{Kotegawa2014,Wu2014,Motizukibook}, with the aim of inducing microcracks that are not completely detrimental to measurements. Indeed, at $p_4$ and $p_5$, which are the runs after $p_3$, RRR decreases to 120 and 96, respectively. Interestingly, although MR(14~T) decreases significantly, the curvature of the MR remains quasilinear. In Fig. \ref{fig2}e, we compare the MR at 3~K and 10~K for $p_2$ and $p_4$, since these pressure values are very close. We find that the MRs at $p_4$ can be satisfactorily scaled to the curves at $p_2$ using the ratio RRR($p_2$)/RRR($p_4$)$\cong$1.62. This exercise works extremely well for the datasets at 3~K.
Therefore, the functional form of the MR is robust in this system and disorder is not expected to play a dominant role.

\begin{figure}[!t]\centering
       \resizebox{8.5cm}{!}{
              \includegraphics{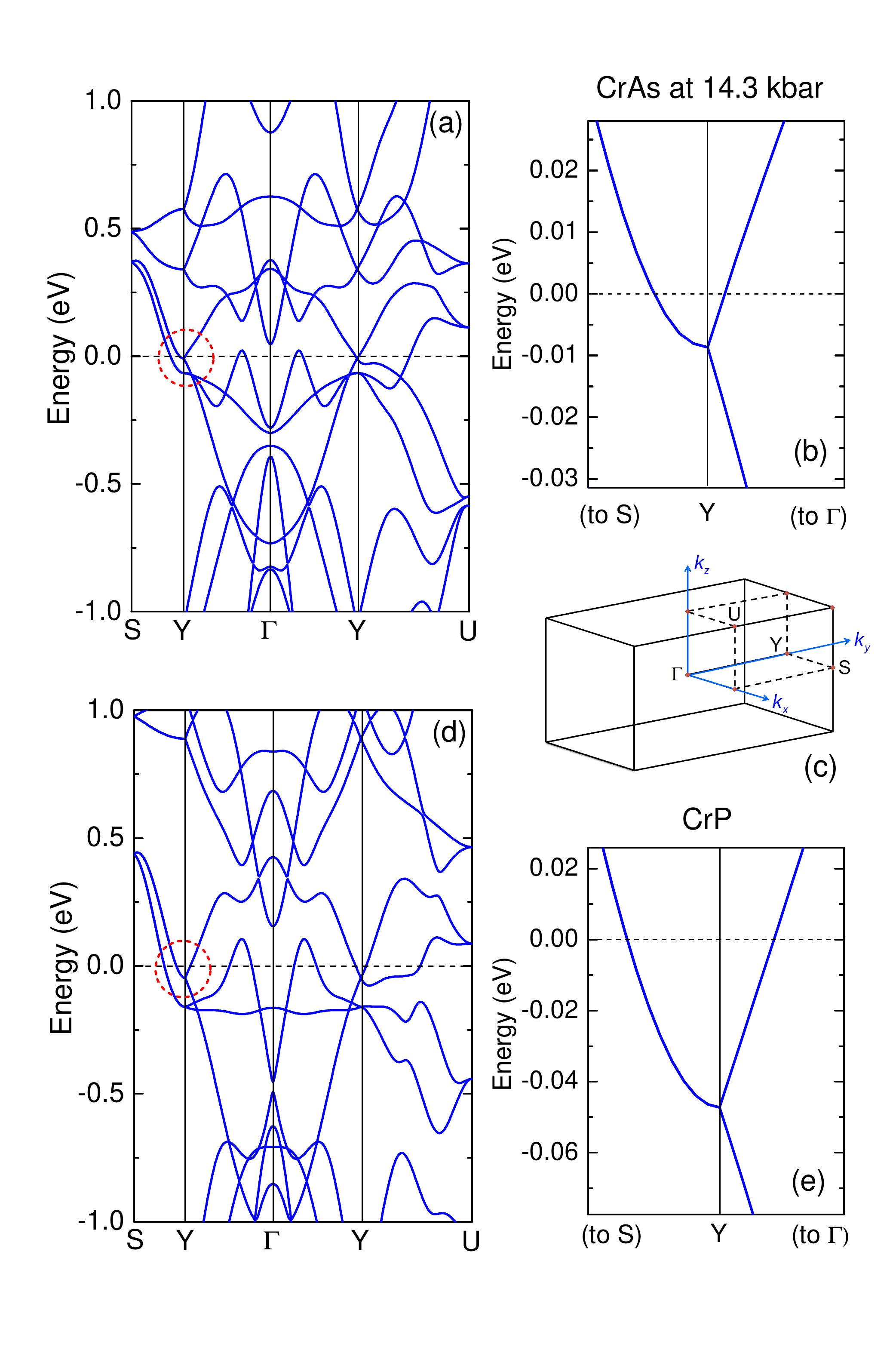}  }           				
              \caption{\label{fig3} {\bf Comparison of the band structure of CrAs at 14.3~kbar and CrP. a,b,} Dispersion relation along several high symmetry directions for CrAs at 14.3~kbar. An expanded view of the band structure near $E_F$ around Y (indicated by the circle) is shown, where a band crossing is found $\sim8.6$~meV below the Fermi energy. The dispersion is linear along Y--$\Gamma$ and parabolic along Y--S. {\bf d,e,} Dispersion relation for CrP. A similar band structure is found near Y, but with a crossing point located at $\sim$47~meV below $E_F$ instead. {\bf c,} The first Brillouin zone of CrAs with the relevant high symmetry points labelled.}
              
\end{figure}
\noindent{\bf Quantum linear MR and extreme quantum limit}\\
According to Abrikosov's theory, a Fermi surface sheet will contribute to a linear quantum MR if all carriers occupy the lowest Landau level \cite{Abrikosov1998,Abrikosov2000}. This is the `extreme quantum limit', and it requires the band extrema to be very near the Fermi energy ($E_F$). Fig \ref{fig3}a shows the band structure of CrAs at 14.3~kbar, calculated using the experimental lattice parameters \cite{Yu2015}. For our multiband system, the band structure near the Y point constitutes a promising case for reaching the extreme quantum limit with a standard laboratory field. At the Y point, a band crossing is found at $\sim$8.6~meV below $E_F$ along the $k_y$ direction, giving rise to a linear energy-momentum dispersion with a slope $\hbar c$. Along the Y--S direction, which is parallel to $\Gamma$--X (Fig \ref{fig3}c), the dispersion is quadratic, parametrized by $m$. The crossing point at Y is thus analogous to the semi-Dirac point discussed in the literature \cite{Dietl2008, Banerjee2009}, except that the parabolic branches are degenerate in our case. Such a hybrid dispersion relation is a result of the nonsymmorphic symmetry of the space group $Pnma$. The two bands crossing at the Y point are degenerate along the $k_x$ direction (Y--S line on the Brillouin zone face), as the nonsymmorphic symmetry protects the degeneracy. Thus, the band crossing has to occur on the Y--S line on which the linear term in $k_x$ is prohibited and the usual quadratic dispersion appears (see Supplementary Information for theoretical proof).  Although a similar band crossing protected by another nonsymmorphic space group $P4/nmm$ was reported in ZrSiS~\cite{Schoop2016}, the crossing point appears 500~meV below $E_F$.  This is to be contrasted with the present case where the crossing point is much closer to $E_F$, and the separation between the crossing point and $E_F$ is tunable by applied pressure (compare Figs. \ref{fig3}b and e). 
The close proximity of the crossing point to $E_F$ implies a small Fermi pocket, which contains neither conventional electrons nor Dirac electrons. In the $k_x$-$k_y$ plane, the dispersion relation around the Y point near $E_F$ can be written as $\varepsilon=\pm\hbar ck_y+\hbar^2k_x^2/(2m)$, where $(k_x, k_y)$ is the wavevector measured from Y. Combining with the quantization condition $S(\varepsilon)=2\pi(n+\gamma)e\mu_0H/\hbar$ within the semiclassical approach, the Landau level spacing is proportional to $H^{2/3}$ when $H\parallel k_z$, and $\gamma\in[0,1]$ is a phase factor which can not be determined from the semiclassical approach. These considerations enable the calculation of the characteristic field $\mu_0H^*$, above which all carriers in this pocket are in the lowest Landau level
\begin{equation}
\mu_0H^*=\frac{8\sqrt{2m}}{6\pi e \hbar c}[f(\gamma)]^{-3/2}((E_F-E_c)+k_BT)^{3/2},
\end{equation}
where $E_c$ is the energy of the crossing point, and $f(\gamma)=(1+\gamma)^{2/3}-\gamma^{2/3}$ is a dimensionless quantity. Since this pocket has an anisotropy $\alpha=k_F^x/k_F^y$ in the $k_x-k_y$ plane, we get another constraint $m$= $\alpha^2(E_F-E_c)/(2c^2)$. 

To scrutinize such a scenario, we extract 
$\mu_0H^*$ above which the MR is linear (Figs. \ref{fig2}f) from our data. The $T$-dependence of $\mu_0H^*$ at $p_2=13.1$~kbar, which is the closest to the pressure value of the calculated band structure, is fitted to $\mu_0H^*=a(T+b)^{3/2}$ (see Supplementary Information for other pressures). $(E_F-E_c)$ can be instantly determined from $b$ to be 3.7~meV. If the band structure displayed in Fig. \ref{fig3}b is rigidly shifted so that $(E_F-E_c)=3.7$~meV, we obtain $\alpha=4.7$. From the fitting parameter $a$, we can estimate that $c$ ranges between $10.5\times10^4$~m/s and $15.6\times10^4$~m/s when $\gamma$ goes from 0 to 1. This is in satisfactory agreement with $c=9.9\times10^4$~m/s extracted from the band structure calculation at 14.3~kbar. Note that the band shifting does not affect $c$, hence the comparison between experimental data and the calculated slope $\hbar c$ is particularly meaningful.

The quantitative agreement between the experimental data and the band structure calculations implies that this particular pocket must dominate the MR. This is extraordinary given the multiband nature of CrAs. Along the $\Gamma$--Y direction, the effective mass is essentially zero because of the linear dispersion. Along the Y--S direction, $m$ ranges between 0.66 -- 0.30~$m_e$ for $\gamma$ between 0 and 1. Therefore, the mobility is high and this pocket dominates the MR. In the vicinity of the magnetic instability, magnetic fluctuations are signicant, as indicated by the enhancement of $A$-coefficient which characterizes the weight of the $T^2$ component in the electrical resistivity \cite{Kotegawa2014,Wu2014}, as well as the spin-lattice relaxation rate \cite{Kotegawa2015}. We speculate that the scattering lifetime is short in other Fermi surface sheets due to strong magnetic fluctuations, and hence they do not contribute significantly to MR.

To further demonstrate the relevance of this pocket, we now examine the bandstructure of CrP, which can again be approximated as a rigid band shift with $(E_F-E_c)$ now equals to $\sim$47~meV (Fig. \ref{fig3}e). If $\gamma$ remains the same in CrP, $\mu_0H^*\approx171$~T at 0~K. Therefore, with a magnetic field of 14~T, it is not possible to have all carriers in the lowest Landau level in CrP. Furthermore, since CrP is far from the magnetic instability, multiple Fermi surface sheets with lower effective mass are contributing to the MR. Hence, the MR is quadratic. \\

\begin{figure}[!t]\centering
       \resizebox{8cm}{!}{
              \includegraphics
              {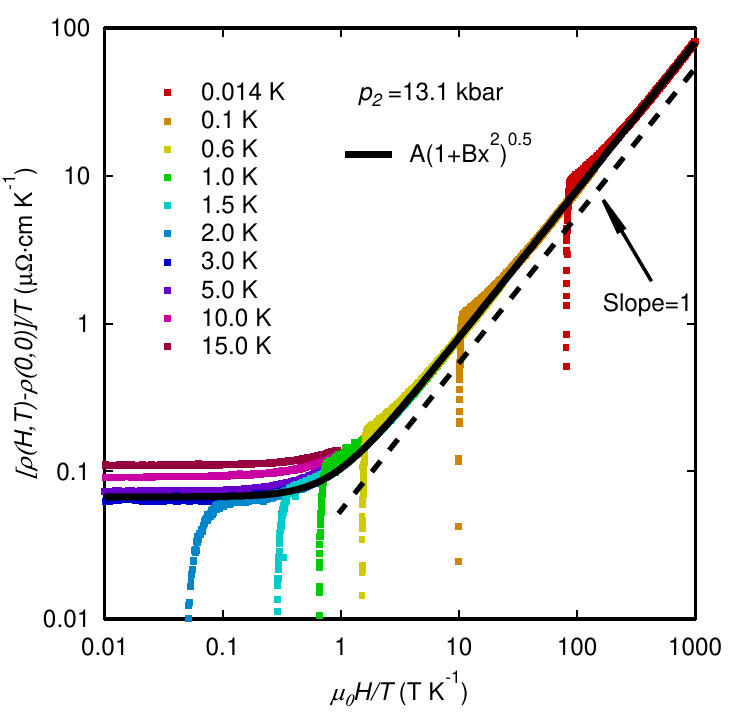}}              				
              \caption{\label{fig4} {\bf $H$-$T$ scaling in CrAs at 13.1 kbar.} The magnetoresistance replotted as $[\rho(H,T)-\rho(0,0)]/T$ against $\mu_0H/T$. The data below 3.0~K can all be described by a single curve $A(1+Bx^2)^{0.5}$ (thick solid curve), which has the same functional form as equation (3). Note that the axes are on logarithmic scales, and the dashed line indicates a unity slope.}
\end{figure}

\noindent{\bf $H-T$ scaling in CrAs under pressure}\\
Recently, quasilinear MR has been observed in BaFe$_2$(As$_{1-x}$P$_x$)$_2$ in the vicinity of the quantum critical concentration $x_c\approx0.3$ \cite{Hayes2016}. Near $x_c$, a linear $T$-dependence of electrical resistivity was reported. Incidentally, the MR is also linear in field, which led to the proposal that the resistivity would be proportional to a new energy scale which is a quadrature sum of the temperature and applied field, \ie, 
\begin{equation}
\rho(H,T)-\rho(0,0)\propto \sqrt{(\alpha k_BT)^2+(\beta \mu_B\mu_0 H)^2},
\end{equation}
or equivalently,
\begin{equation}
[\rho(H,T)-\rho(0,0)]/T \propto \sqrt{1+\lambda\left(\frac{\mu_0H}{T}\right)^2},
\end{equation}
where $\alpha$, $\beta$, and $\lambda$ are numerical parameters. Hence, the resistivity traces plotted as $[\rho(H,T)-\rho(0,0)]/T$ versus $\mu_0H/T$ will all lie on a universal curve. In Fig. \ref{fig4}, we replot our data at $p_2=13.1$~kbar on the axes of $[\rho(H,T)-\rho(0,0)]/T$ and $\mu_0H/T$. All the normal state data below 3~K indeed fall onto the same curve, except for the datapoints just above $H_{c2}$. Note that both axes are on logarithmic scales. Therefore, an excellent agreement is found over {\it five orders of magnitude} in $\mu_0H/T$. Similar behaviour is found for $p_1$ and $p_5$ (see Supplementary Information). These observations imply that the low-temperature $\rho(T)$ in the zero field limit is linear in $T$ over a wide pressure range, even though the system is tuned to a ground state which is far away from magnetism. This is reminiscent of the observation in overdoped cuprates \cite{Cooper2009} and highlights the importance of the proximity to magnetic instability which renormalises the effective masses of the massive bands, thereby allowing a clear dominance of high mobility fermions on magnetotransport properties.  Nontrivial band crossing protected by nonsymmorphic symmetry has been attracting attentions recently \cite{Young2015,Schoop2016,Watanabe2016}; in the case of CrAs, it gives rise to the small Fermi pocket which allows a clear manifestation of the physics at the extreme quantum limit. Such intriguing band structure may also give rise to exotic superconducting properties as recently illustrated in UPt$_3$~\cite{Yanase_UPt3}. Future studies on the influence of the nonsymmorphic symmetry will be an important theme.

\section{Methods}

Single crystals of CrAs were grown with the Sn-flux method from a starting composition of Cr$:$As$:$Sn=1$:$1$:$10 \cite{Kotegawa2014}. Single crystals of CrP were prepared by chemical vapor transport method \cite{Nozue1999}. The powdered polycrystalline sample of CrP was sealed in a silica tube with iodine under vacuum. The charge zone was kept at 900$^{\circ}$C and the growth zone at 800$^{\circ}$C for 2 weeks.

The high pressure environment was provided by a piston-cylinder clamp cell made of MP35N alloys. Glycerin was used as the pressure medium and the pressure values were estimated by the zero field $T_c$ of a Pb manometer placed near the sample. The pressure cell was cooled down to about 15~mK in a dilution refrigerator (BlueFors) equipped with a 14~T superconducting magnet (American Magnetic Inc.) for all pressure points except $p_4=12.5$~kbar, which was examined with a  Physical Property Measurement System (Quantum Design) with a base temperature of 1.8~K. The magnetoresistance was measured with a conventional four-probe method with the magnetic field parallel to the $c$ axis and the current along $a$ axis.

Band structure calculation was performed with the all-electron full-potential linearized augmented plane-wave code WIEN2k \cite{Schwarz2003}. The generalized gradient approximation (GGA) of Perdew, Burke and Ernzerhof (PBE) \cite{Perdew1997} was employed for the exchange-correlation potential. Experimental lattice constants \cite{Yu2015,Selte1975} were used in the calculation and internal structure optimization was performed. The muffin-tin radii were set to 1.96~a.u. for the P atoms and 2.24~a.u. for the Cr and As atoms. R$_{\rm MT}^{\rm min}$K$_{\rm max}$=8 and a $k$-point mesh of 5000 in the first Brillouin zone were used.


\begin{center}
{\bf Acknowledgements}\\
\end{center}
The authors acknowledge stimulating discussion with Anthony Leggett, Xi Dai, Christos Panagopoulos, Takasada Shibauchi and Yasuyuki Nakajima. This work was supported by Research Grant Council of Hong Kong (ECS/24300214), CUHK Startup Grant (No. 4930048), Grant-in-Aids for Scientific Research (No. JP15H03689, JP15H05745, JP15K05885, JP15H05884, JP15K05164, and JP16H00991) from Japan Society for the Promotion of Science, and National Science Foundation China (No. 11504310).

\begin{center}
{\bf Author contributions}\\
\end{center}
S.K.G. conceived and designed the experiment. Q.N., K.Y.Y. and Z.L.L. performed the low temperature measurements. H.K., E.M., H.S. and H.T. prepared the single crystals. W.C.Y. performed the band structure calculations. S.K.G. and Q.N. analysed the data. Y. Y. provided the theoretical support. S.K.G., Y.Y., and Q.N. wrote the manuscript.

\begin{center}
{\bf Competing financial interests}\\
\end{center}
The authors declare no competing financial interests.

\end{document}